\documentstyle[amssymb,prl,aps,twocolumn,epsf,floats]{revtex}
 \newcommand{\PSbox}[3]{\mbox{\rule{0in}{#3}\includegraphics{#1}\hspace{#2}}}

\begin{document}
 \twocolumn [\hsize\textwidth\columnwidth\hsize\csname
   @twocolumnfalse\endcsname
 \title{Microwave photoresponse in the 2D electron
 system caused by intra-Landau level transitions}

 \author{S.~I.~Dorozhkin,$^{1,2}$ J.~H.~Smet,$^1$ V.~Umansky,$^3$ K.~von Klitzing$^1$}
 \address{$^1$Max-Planck-Institut f\"{u}r Festk\"{o}rperforschung,
 Heisenbergstra\ss e 1, D-70569 Stuttgart, Germany}
 \address{$^2$Institute of Solid State Physics, Chernogolovka, Moscow district,
 142432, Russia}
 \address{$^3$Braun Center for Submicron Research,
 Weizmann Institute of Science, Rehovot 76100, Israel}
 \date{\today}

 \maketitle

 \begin{abstract}
 \noindent The influence of microwave radiation on the
 DC-magnetoresistance of 2D-electrons is studied in the regime
 beyond the recently discovered zero resistance states when the
 cyclotron frequency exceeds the radiation frequency. Radiation
 below 30 GHz causes a strong suppression of the resistance over a
 wide magnetic field range, whereas higher frequencies produce a
 non-monotonic behavior in the damping of the Shubnikov-de Haas
 oscillations. These observations are explained by the creation of
 a non-equilibrium electron distribution function by microwave
 induced intra-Landau level transitions.

 \end{abstract}
 \bigskip
  ] 
 \narrowtext The strong recent interest in the microwave
 photoresponse of high-quality two-dimensional electron systems has
 been triggered by the discovery~\cite{mani,zudov2} (also
 ~\cite{dorozh,zudov3,mani1,studenikin,willett}) of zero-resistance
 states in the vicinity of the cyclotron resonance harmonics. These
 states developed out of the minima of earlier reported microwave
 induced magnetoresistance oscillations~\cite{zudov1,ye}. Theories
 capable of accounting for these oscillations are based on several
 different approaches. They include (i) indirect inter-Landau-level
 transitions, which involve the absorption of a microwave quantum
 and are accompanied by scattering processes that alter the
 electron momentum
 ~\cite{ryzhii1,ryzhii2,durst,ryzhii5,ryzhii3,ryzhii4,lei,shikin},
 (more general considerations in terms of the quantum Boltzmann
 equation were given in Ref.~\cite{vavilov}), (ii) the
 establishment of a non-equilibrium electron energy distribution
 function including inverted occupation of the electronic states under
appropriate conditions~\cite{dorozh,dmitriev1,dmitriev2}, (iii)
photon
 assisted quantum tunnelling~\cite{shi}, and (iv) nonparabolicity
 effects~\cite{koulakov}. Most of the cited articles treat the
 magnetoresistance of a homogeneous state and conclude that it may
 become negative. Additional theoretical activity concentrated on
 explaining the appearance of zero-resistance in experiment
 instead. It is argued that negative values of the dissipative
 conductivity lead to instabilities~\cite{zakharov}. As a result,
 an inhomogeneous domain structure, which produces zero resistance,
 may form ~\cite{vavilov,shi,bergeret,andreev,ryzhii6,volkov}.

 Up to now considerations were limited mainly to the weak magnetic
 field range where the microwave frequency $\omega$ exceeds the
 cyclotron frequency $\omega_{\rm c}$ and inter-Landau level
 transitions are of great importance. Here, we are concerned with
 the influence of the microwave radiation on the magnetoresistance
 and the amplitude of the Shubnikov-de Haas (SdH) oscillations in
 the opposite regime when $\omega<\omega_{\rm c}$.
Inter-Landau-level transitions then no
 longer play a role. At comparatively low-frequency
 radiation, we experimentally observe a strong
 suppression of the magnetoresistance accompanied by a drop in the
 amplitude of the SdH oscillations~\cite{comment}. At higher radiation
 frequencies, the SdH oscillations are also strongly damped except
 for a narrow region of the magnetic field $B$, where the amplitude is
 rather insensitive to the radiation. We attribute the
 magnetoresistance and the SdH oscillation suppression to the
 non-equilibrium electron occupation caused by microwave induced
 intra-Landau level transitions. The unusual non-monotonic damping
 of the SdH-oscillations at higher frequencies reflects the
 crossover from the inter- to intra-Landau-level transition regime
 in the case of non-overlapping levels.

 We have measured 3 Hall bar samples produced from two different
  wafers of the standard architecture containing a
 two-dimensional electron system at a single remotely doped
 GaAs/AlGaAs heterojunction with a spacer width of 80 nm. The
 mobility after illumination reached
 values between $6-15 \times 10^{6}\ {\rm cm}^2/{\rm V\,s}$ at a
 typical saturated density of $3\times 10^{11}\ {\rm
 cm}^{-2}$. The channel widths of the samples were equal to
 0.05, 0.2 and 0.4 mm. All samples demonstrated
 qualitatively the same results. The main difference was the
 amplitude of the microwave induced oscillations.
 The samples were mounted in a waveguide with a
 cross-section of $16\times 8\ \mbox{mm}^2$ (WR28) and submerged in
 pumped $^3{\rm He}$. The $\rho_{\rm xx}$ and $\rho_{\rm xy}$
  magnetoresistivity tensor components were measured with lock-in technique using a 10 Hz sinusoidal
 current.

 \begin{figure}[tbp!]
 \PSbox{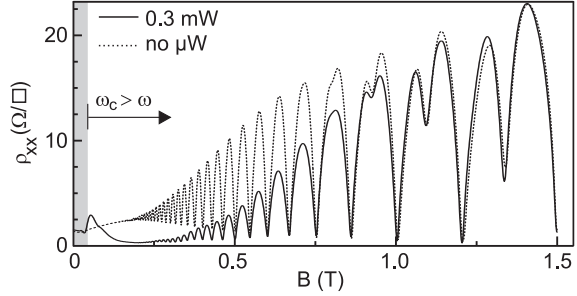}{8cm}{4.5cm} \caption{Magnetoresistivity
 $\rho_{\rm xx}$ versus $B$ in the absence of
  radiation (dotted line) and under 17 GHz radiation
 (solid line) at T=0.4 K and density $n_{\rm s}=2.92 \times
 10^{11} {\rm cm}^{-2}$. The microwave power $P=0.3\,{\rm mW}$ was
 measured at the oscillator output. The microwave electric field is
 perpendicular to the current.} \label{fig1}
 \end{figure}
 Typical experimental curves are shown in Fig.~1. In the absence of
 microwave radiation, $\rho_{\rm xx}$ exhibits the usual SdH
 oscillations. Radiation with frequency $\omega$ dampens
 these oscillations at low $B$-fields and gives rise to the
 microwave induced magnetoresistance oscillations (see also Fig. 2)
 with minima located where $\omega \approx \omega _{\rm c} (k +
 1/4)$ (Here $k=1,2,\ldots$). As seen in Fig.~1 low frequency
 radiation (for our samples less than 40 GHz) also dramatically
 suppresses the average magnetoresistivity within a wide
 $B$-field range and even when $\omega_{\rm c}\gg \omega$. For the
 particular case of Fig.~1, the magnetoresistivity $\rho_{\rm xx}$  at for instance
 0.2 T ($\omega_{\rm c}/\omega \approx 5$) is reduced by one order
 of magnitude and becomes less than the zero-field
 resistivity by about a factor of 5. In the single-mode regime of
 the waveguide, i.e. for frequencies below 19 GHz, this drop in $\rho_{\rm xx}$
  was observed for both orientations (parallel
 and perpendicular) of the microwave electric field with respect to
 the excitation current. Fig.~2 depicts how this $\rho_{\rm xx}$
 suppression evolves with microwave frequency~\cite{sdh}. It
 reduces at higher frequencies and disappears near 40-50 GHz. At these higher
 frequencies, plotting the magnitude of each SdH oscillation,
 $A(P)$,
 normalized to its amplitude in the absence of microwave
 radiation, $A(0)$,
 reveals a non-monotonic behavior as a function of $B$
 as seen for instance in Fig.~3b. Note
 that data presented in Fig.~3 have been measured at more than one
 order of magnitude smaller power than data depicted in Fig.~2. Fig.~3a shows
 such low power data for 40 GHz. Maxima of $A(P)/A(0)$ in Fig.~3b, nearly symmetrically
 arranged around $B=0$, are located in the vicinity of the
 second cyclotron resonance subharmonic, i.e. when
 $\omega=\omega_{\rm c}/2$. The asymmetry of the curves (especially
 at lower frequency) with respect to $B = 0$ is tentatively
 assigned to the excitation of edge magnetoplasmons by the
 radiation in view of their chiral properties~\cite{kukush}.
 At the maxima, $A(P)/A(0)\approx 1$, i.e.~the SdH oscillation amplitude
 is
 insensitive to the applied low power microwave radiation. The red
 shaded box in Fig.~3a highlights this region for the 40 GHz data.
 \begin{figure}[tbh!]
 \PSbox{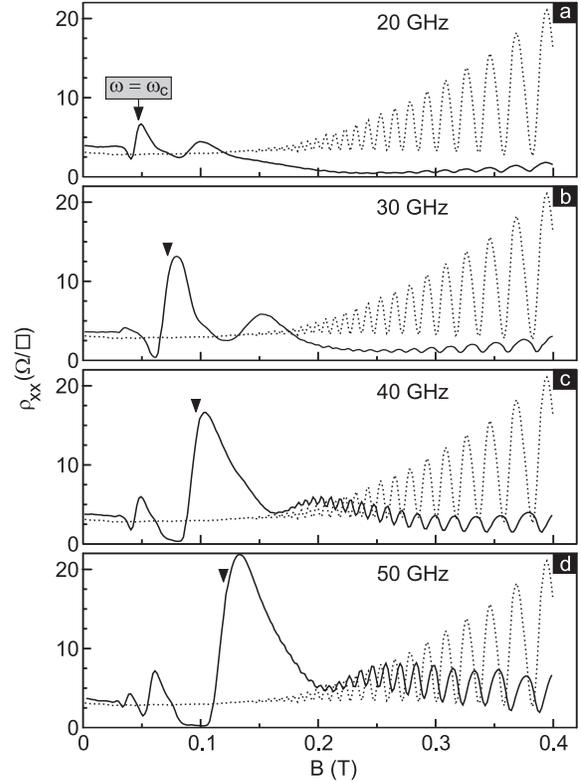  voffset=-20}{8cm}{11cm}
 \caption{Magnetoresistivity $\rho_{\rm xx}$ versus
 $B$ without radiation (dotted lines) and under
 microwave radiation (solid lines) for the marked frequencies
  at T=0.4 K and $n_{\rm s}=2.8 \times 10^{11} {\rm
 cm}^{-2}$. The oscillator output power $P$ was equal to $2\ {\rm
 mW}$. The positions of the cyclotron resonance are marked by
 arrows.}
 \end{figure}
 \begin{figure}[tbh!]
 \PSbox{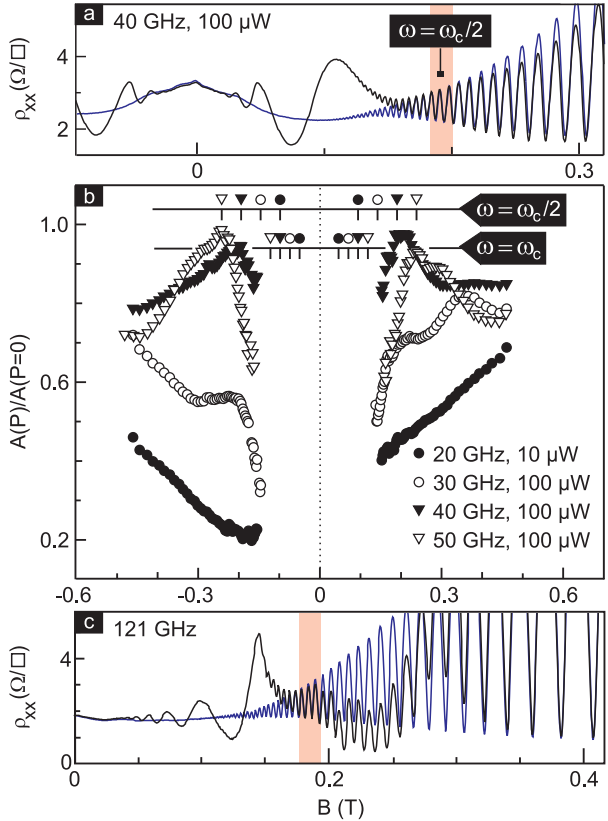 voffset=-20}{8cm}{11.6cm} \caption{
(a) $\rho_{\rm xx}$ versus B without (blue) and with $100\ {\rm
\mu W}$ of 40 GHz radiation. (b)
 The
 amplitude of each SdH-oscillation  under microwave
 radiation with power $P$, $A(P)$, normalized to its dark value,
 $A(P=0)$,
 is plotted for different microwave frequencies. The location of
 the cyclotron resonances and second cyclotron resonance
 subharmonics are marked. (c) $\rho_{\rm xx}$ vs.~B without (blue) and with 121 GHz radiation. The red shaded
 box,
 where $\omega \approx 3 \omega_{\rm c}/2$, demarcates as in (a) the region where SdH oscillations do not respond to microwaves.}
 \end{figure}

 To account for these observations, we first address qualitatively
 what microwave absorption processes can take place when
 considering energy conservation, while assuming other selection
 rules are relaxed due to the inevitable disorder in the samples.
 In Fig.~4, the regions in the ($\omega_{\rm c},\omega$)-plane
 where inter- and intra-Landau level transitions can occur are
 color coded. These areas are bounded by two lines:
 $\omega=\omega_{\rm c}-2\Gamma/\hbar$ and $\omega=2\Gamma/\hbar$.
 Here, $\Gamma$ is half the width of a broadened Landau level. When
 the cyclotron radius $r_{\rm c}$ exceeds the characteristic length
 scale, $\lambda$, of the random potential, this width increases
 with the square root of the applied $B$~\cite{ando,raikh}. The border lines for inter- and
 intra-Landau level transitions then intersect when $\omega_{\rm
 c}=\omega_{\rm c0}$ and $\omega=\omega_0=\omega_{\rm c0}/2$, i.e.
 when the microwave frequency coincides with the second subharmonic
 of the cyclotron resonance frequency $\omega_{\rm c0}$.
 For $\omega<\omega_0$, there are always
 transitions possible. They may
 modify the electron distribution function, which affects
 the conductivity
 (resistivity) and the SdH oscillation amplitude~\cite{dorozh,dmitriev1,dmitriev2} and produces the dramatic drop
 in $\rho_{\rm xx}$ as shown later.
 However when scanning the $B$-field at a fixed microwave
 frequency slightly exceeding $\omega_0$, the white region where we
 anticipate very weak absorption is briefly entered.
It is likely responsible for the maxima in the 40 and 50 GHz
 curves of Fig.~3b where the amplitude of the SdH oscillations is
 only weakly influenced by the radiation.
A similar non-monotonic behavior of
 the SdH oscillation amplitudes is expected from Fig.~4 at $\omega_{\rm c}=\omega_{\rm c0}$
and $\omega=3 \omega_{0}$. This was indeed confirmed in
experiment. For $f = 121\ {\rm GHz}$ in Fig.~3c near $B\approx
0.18\ {\rm T}$, the SdH oscillations are insensitive to radiation.
 Surprisingly, such
 straightforward considerations explain quantitatively the
 experimentally observed position of the maxima. It is worth noting
 that, within this picture, the decay of $A(P)/A(0)$ beyond the
 maxima in Fig.~3b at higher absolute values of $B$ implies that the
 Landau level width increases with $B$.  Then, for frequencies
 exceeding $\omega_{\rm 0}$ the low field region with predominant
 inter-Landau level transitions is followed by no transitions and
 finally by a field regime where intra-Landau-level transitions
 occur. This field dependent broadening is distinctive of a
 short-range random potential for which $\lambda \leq r_{\rm
 c}$~\cite{ando,raikh}. In remotely doped heterojunctions, the
 shortest range of the random potential is frequently given by the
 spacer width. In our samples the cyclotron radius becomes smaller
 than the spacer width at $B > 1\ {\rm  T}$. The presence of
 fluctuations on a length scale $\lambda < r_{\rm c}$ at low fields
 allows intra-level transitions.
 \begin{figure}[tbh!]
 \PSbox{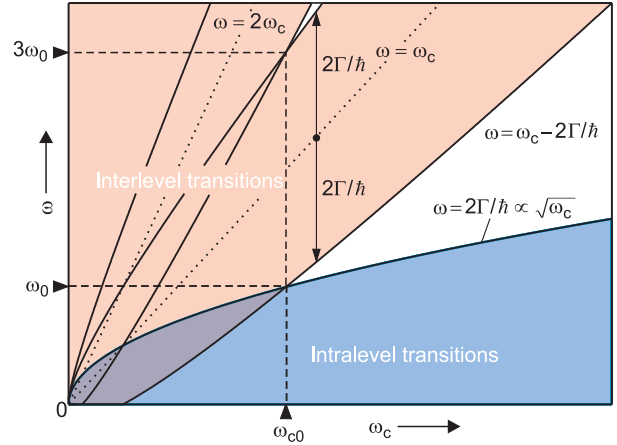}{8cm}{6.4cm} \caption{The
 ($\omega_{\rm c},\omega$)-plane. Regions where inter- and
 intra-Landau level transitions can occur under microwave radiation
 are color coded. The main boundaries are formed by
 $\omega=\omega_{\rm c}-2\Gamma/\hbar$ and $\omega= 2\Gamma/\hbar$.
 They intersect at $\omega=\omega_0=\omega_{\rm c0}/2$. In the
 white regions energy conservation prohibits transitions. The
 position of the cyclotron resonance and its second harmonic
 (straight dotted lines) are also shown.}
 \end{figure}

To substantiate the assertion that the strong reduction of the
average magnetoresistance for $\omega\ll\omega_{\rm c}$
 can be attributed to changes in the electron distribution function, we have analyzed
 whether the
 recently proposed theory in Ref.~\cite{dmitriev2} is
 capable of
 reproducing this phenomenon. As initially suggested in
 Ref.~\cite{dorozh},
this theory
 considers the nonequilibrium population of
 electronic states. In addition, it allows for finite temperature
 and inelastic relaxation. We have solved numerically the
 equation for the non-equilibrium distribution function of
 Ref.~\cite{dmitriev2} (Eq. (2)) for the case of non-overlapping
 Landau levels, while omitting the term describing
 effects caused by the dc electric field. We further assume that
 spin splitting is not resolved.
 Calculated traces of $\rho_{\rm xx}$ for frequencies
  $\omega_1<\omega_0$ and $\omega_2>\omega_0$ are shown in Fig. 5.
 A damping of the SdH oscillations at intermediate values of
 $\omega_{\rm c}/\omega$ accompanied by a strong suppression of the
 average magnetoresistivity is apparent in Fig.~5a for
 $\omega_1<\omega_0$, qualitatively confirming observations in
 Fig.~1 and 2.
The magnetoresistance
 suppression is a result of the strong modification of the
 distribution function mainly within the highest occupied Landau
 level caused by intralevel transitions as evident from the right inset.
 Even in the presence of inelastic relaxation, energy ranges with
 inverted electron populations exist, which yield a negative
 contribution to the magnetoresistance not unlike what occurs at
 $\omega > \omega_{\rm c}$ where the zero resistance states
 develop.
 Fig. 5b for $\omega_2>\omega_0$
 demonstrates the insensitivity of the SdH oscillations located
 around the second cyclotron resonance subharmonic (i.e. when
 $\omega = \omega_{\rm c}/2$) to the microwave power in a narrow
 magnetic field region, highlighted in red. This is consistent
 with the data in Fig.~3.

 From the appearance of the
 maxima in $A(P)/A(0)$ at 40 GHz and their location in Fig.~3b, the Landau level
 width can be estimated. The data suggest $2\Gamma/\hbar\omega_{\rm
 c}=0.5$ when $B\approx 0.18\,{\rm T}$. This estimate seems to be
 in reasonable agreement with the lower magnetic field limit where
 we are able to resolve the SdH oscillations (0.1 T at T=0.4 K).
\begin{figure}[tbh!]
 \PSbox{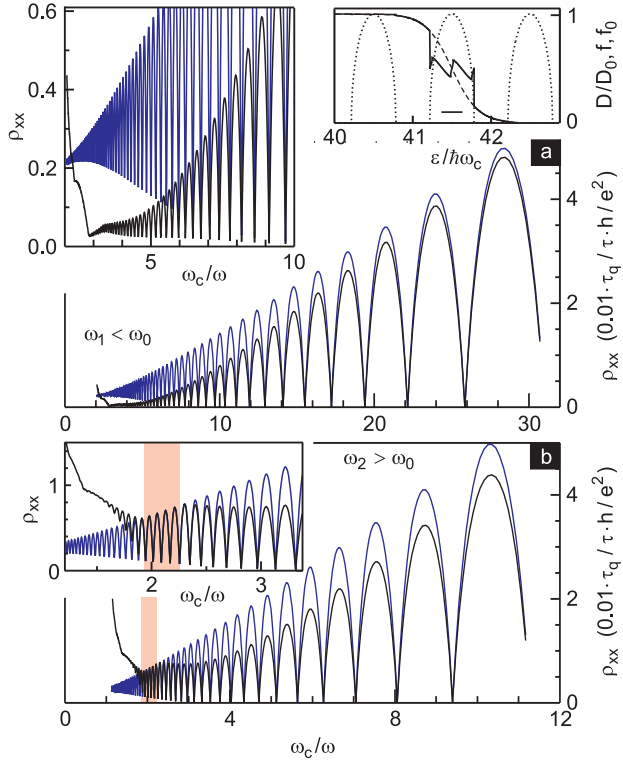}{8cm}{10.8cm} \caption{Calculation of
 $\rho_{\rm xx}$ vs.~$\omega_{\rm c}/\omega$ (or B) (black solid
 lines) for frequencies $\hbar\omega_1 \approx 0.42
 \hbar \omega_{\rm 0} \approx 6.5\times 10^{-3}\epsilon_{\rm F}$
 (a) and $\hbar\omega_2= 1.15 \hbar\omega_{\rm 0} \approx
 0.018\epsilon_{\rm F}$ (b) and for the  dimensionless microwave
 powers $P_{\rm \omega}^{\rm (0)}$ {\protect \cite{dmitriev2}}
 (proportional to $\omega^{-4}$) of 100 and 10 respectively. The
 temperature $T\approx 4.0\times 10^{-3}\epsilon_{\rm F}$. Here
 $\epsilon_{\rm F}$ is the Fermi energy at zero $B$-field.
 Blue lines present the calculated magnetoresistivity in the
 absence of radiation. The left insets are expanded views (same units as main panels). The right inset in
 the top panel depicts the
 semi-elliptical density of states $D/D_0$ (dotted line){\protect
 \cite{ando}} and calculated distribution function $f(\epsilon)$
 (solid line) under microwave radiation for the point $\omega_{\rm
 c}/\omega\approx 3.7$, corresponding to filling factor 83. Here,
 $f_0(\epsilon)$ (dashed line) is the Fermi distribution function,
 with $D_0=2/\pi^2l_{\rm B}^2\Gamma$ and  $l_{\rm B}$ the magnetic
 length. The value $\omega_1/\omega_{\rm c}$ for the chosen
 magnetic field is plotted in the inset as a line segment. In the
 bottom panel, the magnetic field interval where the SdH
 oscillations are not affected by radiation is demarcated by the
 red shaded box.}
 \end{figure}

 In summary, we have presented experimental data showing a dramatic
 suppression of the magnetoresistance across a wide field range
 where microwaves can, under the condition of single photon
 absorption, only induce intra-Landau level transitions. The
 regime where intra-Landau-level transitions start to take over from
 inter-Landau level transitions is detected from the anomalous
 damping behavior of the SdH oscillations. This anomaly allows to
 estimate the Landau level width.

 The authors gratefully acknowledge fruitful discussions of the
 manuscript with A.~Mirlin, D.~Polyakov, and I.~Dmitriev
 and financial support by
 INTAS, RFBR (SID), the GIF and the DFG.


\begin{references}
\bibitem{comment} The suppression of $\rho_{\rm xx}$ can also be
 found in some low microwave frequency data of
 Ref.\cite{dorozh,willett,ye}.
\bibitem{sdh} High power microwaves may induce a small shift
 of the SdH-oscillation minima as in Fig.~2.
\end{references}
 \end{document}